\documentstyle[twoside,fleqn,espcrc2,psfig,graphics]{article}
\newcommand{\AmS}{{\protect\the\textfont2
  A\kern-.1667em\lower.5ex\hbox{M}\kern-.125emS}}

\title{ Monopoles, Vortices and Confinement in SU(3) Lattice Gauge 
Theory}
\input epsf
\author{Roy Wensley \address{Department of Physics and Astronomy, 
Saint Mary's College, Moraga, California, 94575, USA} and John Stack 
\address{Department of Physics, University of Illinois at 
Urbana-Champaign, 1110 West Green Street, Urbana, Illinois, 68101}
\thanks{This work was supported in part by the National Science 
Foundation, NPACI, and the Saint Mary's College 
Faculty Development Fund.  }}
       
\begin{document}

\begin{abstract}
We present results for the heavy quark potential computed in SU(3) 
from magnetic monopoles and from center vortices.  The monopoles are 
identified after fixing SU(3) lattice configurations to the maximal 
abelian gauge.  The center vortices are identified after using an 
indirect center gauge fixing scheme which we describe for SU(3).  Z(3) 
center vortices are extracted and used to compute the potential.  The 
values of the string tensions from monopoles and vortices are compared 
to the full SU(3) string tension.
\end{abstract}

\maketitle

A clear understanding of the confinement mechanism in QCD is still 
elusive.  Two particular mechanisms have gained the most attention: 
magnetic monopoles and center vortices.  Here we present results of 
calculations in SU(3) pure gauge theory of the heavy quark potential 
using methods which test these two different mechanisms.

\vspace{.25in}

\section{Monopoles and the Maximal Abelian Gauge}

The maximal abelian gauge is obtained in $SU(3)$ gauge theory by
maximizing the lattice functional
\begin{eqnarray*}
S= \sum_{x,\mu}      & & \!  \!  \!  \! \! \! \! \! \! \! \! \!
{1 \over 2}  [Tr(U_{\mu}(x)\lambda_{3}
U^{\dagger}_{\mu}(x)\lambda_{3})\\
 & + &  Tr(U_{\mu}(x)\lambda_{8} 
 U^{\dagger}_{\mu}(x)\lambda_{8})],
\end{eqnarray*}
where $\lambda_{8}$ and $\lambda_{3}$ are the diagonal generators of 
the fundamental representation of $SU(3)$.  The maximization is done 
by performing local $SU(3)$ gauge transformations on the links, which 
in practice is done using an $SU(2)$ subgroup update\cite{cab}.

Once the maximal abelian gauge is obtained, two independent abelian 
gauge fields are projected out, leaving $U(1)\times U(1)$ 
gauge fields.  Magnetic monopoles are then identified in the abelian 
configurations.  The lattice monopoles are used to directly 
compute the heavy quark potential\cite{banks,stack,yosh}.

\vspace{.25in}

\section{Indirect Center Gauge}

The indirect center gauge was used in the first lattice studies of 
projected center vortices in $SU(2)$\cite{deb}.  The indirect center 
gauge is obtained in the following way.  A non-abelian lattice gauge 
configuration is fixed to the maximal abelian gauge.  Abelian 
projection is performed obtaining the abelian subgroup configurations: 
In $SU(2)$ this procedure obtains $U(1)$ gauge fields, while in 
$SU(N)$ it yields $[U(1)]^{N-1}$ gauge fields.  The remaining $U(1)$ 
configurations are then gauge fixed so that the gauge fields are 
brought as close as possible to the elements of the center group of 
the original $SU(N)$ theory.  For the case of $SU(2)$ this corresponds 
to maximizing
$$R_{2}=\sum_{x,\mu}\cos^{2}\left( \theta (x,\mu) \right),$$
where $\theta(x,\mu)$ parameterizes the abelian links.  The 
maximization is obtained by iteratively performing $U(1)$ gauge 
transformations.  The final step to extract the vortices involves 
another projection of the lattice gauge fields onto the group elements of 
$Z(2)\in\{-1,+1\}\times {\bf 1},$ where $\bf 1$ is the unit matrix.

In $SU(3)$ after maximal abelian gauge fixing and projection is done, 
the links are elements of $U(1) \times U(1)$.  The 
$U(1) \times U(1)$ links can be parameterized by the matrix: 
$$\!  \!  \!  \! \! \! \! \! \! \! \! \! U_{\mu}(x) = $$
$${\rm diag}(e^{i\theta_{1}(x,\mu)},
e^{i\theta_{2}(x,\mu)},
e^{-i(\theta_{2}(x,\mu)+\theta_{1}(x,\mu))}). $$ 
$U(1)\times U(1)$ gauge transformations are then applied to these 
links to bring them as close as possible to the elements of the $SU(3)$
center group, 

$$Z(3) \in \{ \exp(-i 2\pi/3), 1, \exp(+i2\pi/3)   \}\times {\bf 1}.$$ 

The necessary gauge 
transformations can be parameterized 
using the fundamental, diagonal generators of $SU(3)$:

$$g(x)=e^{i\tilde{\lambda}_{8}\alpha_{2}(x)} e^{i\lambda_{3}\alpha_{1}(x)},$$

where some numerical factors are absorbed into the $\alpha$'s and we 
use
$$\tilde{\lambda_{8}}={\rm diag}(1,1,-2)$$
$$\lambda_{3}={\rm diag}(1,-1,0).$$

The center gauge in $SU(N)$ is obtained by maximizing the quantity
$$R_{N}=\sum_{x,\mu} \left|{\rm Tr}U_{\mu}(x)\right|^{2}.$$
For the indirect case in $SU(3)$, this is accomplished by applying the 
$U(1)\times U(1)$ gauge transformations described above to the 
diagonal $U(1) \times U(1)$ links to maximize the quantity
\begin{eqnarray*}
R_{3}=\sum_{x,\mu}& &\!\!\!\!\!\!\!\! 3 + 
2\cos(\theta_{1}(x,\mu)-\theta_{2}(x,\mu))\\
 &  &+\cos(\theta_{1}(x,\mu)+2\theta_{2}(x,\mu))\\
 &  &+\cos(\theta_{2}(x,\mu)+2\theta_{1}(x,\mu)). \label{R}
\end{eqnarray*}
Gauge transformations must be applied iteratively to each site 
independently to maximize the global quantity.  
There are two local gauge conditions that must be met when the center 
gauge has been obtained.  The first condition comes from the 
$\tilde{\lambda_{8}}$ contribution:
\begin{eqnarray*}
& \sum_{\mu}  \partial^{-}_{\mu} \left[\sin(2\theta_{1}(x,\mu) 
+\theta_{2}(x,\mu))\right.&\\
&\left.   + \sin(2\theta_{2}(x,\mu) + \theta_{1}(x,\mu))\right]&=0,\\
\end{eqnarray*}
where $\partial^{-}_{\mu}$ is the lattice derivative
$$\partial^{-}_{\mu} f(x) = f(x)-f(x-\hat{\mu}).$$
The second gauge condition to be met comes from the $\lambda_{3}$ 
contribution:
\begin{eqnarray*}
&\sum_{\mu} \partial^{-}_{\mu} 
 [2\sin(\theta_{1}(x,\mu)-\theta_{2}(x,\mu))&\\
&+\sin(2\theta_{1}(x,\mu)+\theta_{2}(x,\mu))&\\
&-\sin(2\theta_{2}(x,\mu)+\theta_{1}(x,\mu))]&=0.
\end{eqnarray*}
Numerically, neither term can actually be zero.  In practice, 
numerical iterations are continued until both terms are on the order 
of $10^{-7}$.  In the continuum these two gauge conditions are 
equivalent to the  Landau gauge:

$$\partial_{\mu} \theta_{1}(x,\mu) =0$$
$$\partial_{\mu} \theta_{2}(x,\mu) =0$$

\section {Calculations}

We generated 400 equilibrium $SU(3)$ lattice configurations on a 
$10^{3} \times 16$ lattice with $\beta = 5.9$.  At this value of 
$\beta$ the correlation length is about 4 lattice spacings.  Using 
these configurations and smearing techniques\cite{ape} we computed the 
heavy quark potential and measured the full $SU(3)$ string tension, 
$\sigma_{f}$.  We found $\sigma_{f}=.068(3)$.

Next, the 400 configurations were fixed to the maximal abelian gauge 
and the two species of abelian fields\cite{yee94} were projected out.  
Overrelaxation was used with $\omega=1.7$\cite{mandula}.  In each 
abelian configuration Dirac strings were located using the $1 \times 
1$ plaquettes and the two species of monopole configurations were 
obtained.  The monopole configurations were used to compute the heavy 
quark potential (see Fig.~(2) below).  Results for both monopole 
species were found to be the same.  The string tension from the 
potential computed using monopoles was found to be 
$\sigma_{mon}=.047(2)$.  This is about $30\%$ lower
than the full $SU(3)$ string tension $\sigma_{f}$.  We believe this is 
the first quantitative calculation of the SU(3) string tension from 
monopoles, although hints that the monopole string tension is low have 
appeared before\cite{yosh}.

After gathering the abelian configurations, we next applied our 
abelian center gauge fixing algorithm.  No closed form solution of the 
parameters $\alpha_{1}$ and $\alpha_{2}$ could be found for the 
transcendental equations resulting from maximizing $R_{3}$.  The 
$\alpha$'s were determined iteratively.  The value of the normalized 
abelian trace computed from using the $U(1) \times U(1)$ links was 
observed to change from an average value of 0.525 before center gauge 
fixing to 0.922 after gauge fixing.

\begin{figure}[htb]
\vspace{9pt}
%\hspace*{-2.5in}
\psfig{file=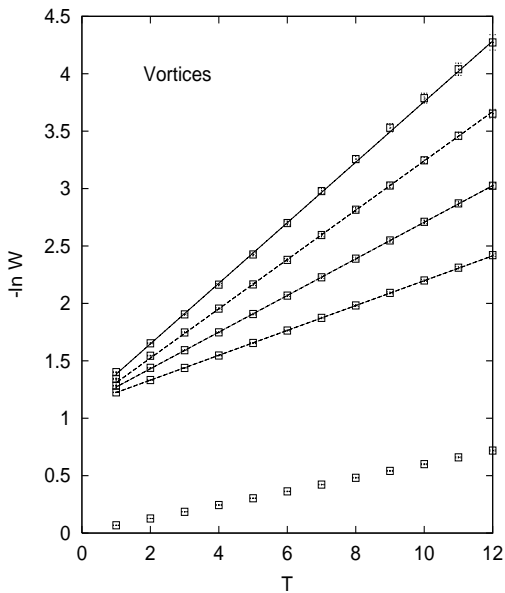,bbllx=1cm,width=7cm,height=8.cm}
%,angle=-90}
\caption{Wilson loops computed using $Z(3)$ links obtained after
indirect center projection.  The loop values are for $R=1$ to
$R=5$ ordered from bottom to top, respectively}
\label{fig1}
\end{figure}

From the center gauge fixed configurations, the links were replaced by 
the $Z(3)$ phase each was closest to.  Wilson loops were computed 
using the $Z(3)$ configurations.  The Wilson loop data is presented in 
Figure~(1).  Fits to the potential computed using the $Z(3)$ Wilson 
loops gave a string tension of $\sigma_{vort}=.060(4)$. 
Thus, we find the string tension from indirect center projection to be 
too low.

\begin{figure}[htb]
\vspace{9pt}
%\hspace*{-2.5in}
\psfig{file=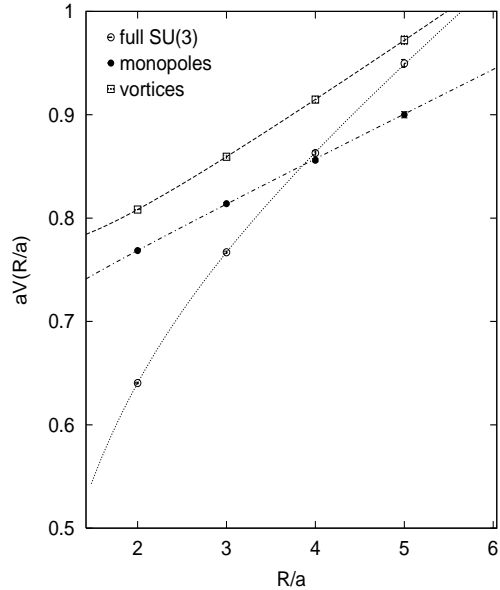,bbllx=1cm,width=7cm,height=8.cm}
%,angle=-90}
\caption{Comparison of the three heavy quark potentials, computed 
directly from the $SU(3)$ gauge fields, using monopoles, and 
using projected center vortices.}
\label{fig2}
\end{figure}

In Figure~(2) we compare the heavy quark potentials.  The potentials 
computed from the monopoles and projected center vortices show little 
Coulomb behavior at small $R$ as compared to the full $SU(3)$ 
potential.  This has been observed in other calculations in $SU(2)$ 
and $U(1)$\cite{stack,deb,stack92}.  Calculations which demonstrate 
the confinement mechanism are not required to reproduce the short 
distance behavior of the potential, only the string tension.

\section{Conclusions}

We have presented results for the string tension computed using 
monopoles and center vortices.  The calculations use operations 
defined at the level of a single lattice spacing.  Neither monopoles 
nor center vortices reproduce the full $SU(3)$ string tension.  There 
is still a need to refine the methods used here if one is to discard 
or accept monopoles or center vortices as the confinement mechanism in 
QCD. We conclude by discussing possible reasons to suspect systematic 
errors exist in these calculations.

Monopoles and vortices are computed using objects defined on the 
single link level.  The results for the string tension may be 
corrupted by unphysical degrees of freedom due to misidentification of 
monopoles and center vortices by using operators defined at the single 
link level.  Also, we may not be using volumes large enough to 
suppress finite volume effects since monopoles and center vortices are 
objects that likely extend over many lattice spacings.  To investigate 
the first problem, we need to construct other methods for identifying 
monopoles and center vortices.  For the second problem we need to work 
on larger lattices.  Using larger lattices is difficult due to the 
computationally intensive nature of the gauge fixing methods.

The numerical procedures for the maximal abelian and center gauge 
fixing suffer from Gribov copy problems.  In particular, for the 
direct and indirect center gauge fixing procedures in $SU(2)$ the 
effects of Gribov copies on the numerical value of the string tension 
are large\cite{born}.  

The effect of taking account of the Gribov
ambiguity for $SU(2)$ for all cases investigated so far, is to
decrease the corresponding string tension.  So for monopoles in the 
maximal abelian gauge, the monopole string tension for one gauge 
fix/configuration is quite close to the full $SU(2)$ value \cite{stack}, 
and decreases
by $O(10\%)$ when account is taken of the Gribov ambiguity by generating many
copies/configuration \cite{bali}.  Our result for the 
$SU(3)$ monopole string tension
is considerably lower than the full $SU(3)$ value even at one gauge 
fix/configuration, and would presumably be reduced somewhat more when gauge
ambiguities are accounted for.   Thus the $SU(2)$ and $SU(3)$ cases 
appear to be rather different.
	Given the large noise level of our data, we did not succeed in
extracting a string tension at the $U(1)\times U(1)$ level, before 
projecting to monopoles or vortices.  Given our low results for the monopole
and center vortex string tensions, this is now an important quantity to measure.
A failure of abelian projection at the $U(1) \times U(1)$  level for $SU(3)$
would have important implications for ideas on confinement in $SU(3)$.  We
plan to return to this question in future work.

\end{document}